
\leftskip=0pt \parskip=0pt \parindent=0pt
\tolerance 10000

\font\twelverm=cmr10 scaled 1200
\font\fourteenbf=cmbx10 scaled 1440
\nopagenumbers
\headline={\tenrm\hfil\folio}
\vskip .25in
\baselineskip=18pt
\def\p#1 #2 {{{\partial #1}\over {\partial #2}}}
\def\ov#1 #2 {{{#1}\over {#2}}}
\vskip .5in
\centerline {\fourteenbf ANALYTIC SOLUTIONS OF THE VECTOR BURGERS' EQUATION}
\vskip .2in

\centerline {\twelverm Steven Nerney$^{1}$, Edward J. Schmahl$^{2}$
, and Z. E. Musielak$^{3}$}
$^1$ National Research Council Associate, NASA-Marshall Space Flight
Center, Alabama 35812.

$^2$ Astronomy Department, University of Maryland, College Park,
MD 20742.

$^3$ Department of Mechanical and Aerospace Engineering,
and Center for Space Plasmas and Aeronomic Research, University
of Alabama at Huntsville, Huntsville, Alabama 35899.\hfill
\bigskip
Accepted by the Quarterly of Applied Mathematics\hfill
\bigskip
September 3, 1993
\vskip .4in
{\leftskip = 1in \rightskip =1in
{\bf Abstract.}
 The well-known analytical solution of Burgers' equation is extended
to curvilinear coordinate systems in three-dimensions by a method
which is much simpler and more suitable to practical applications
than that previously used [22]. The results obtained are applied
to incompressible flow with cylindrical symmetry, and also to the
decay of an initially linearly increasing wind. }
\medskip

\vskip .4in
{\bf 1. Introduction}
\vskip .1in
Burgers' equation is a well-known example of a
non-linear partial differential equation whose solution can be
constructed
from a linear partial differential equation. It is, to the best
of our knowledge, the only such example. Much of the interest
in this equation arises because it is a very simple form of the
Navier-Stokes equation in the one-dimensional, cartesian,
time-dependent,
compressible, viscous limit [17]. An early study derived two
steady-state solutions [1],
and then the equation was used in studying the decay of free
turbulence [4,5]. The importance of the equation is due to the
non-linear term, $u\p  u x $, which allows the calculation of the
modification of the velocity due to the exchange of momentum
between a variety of different length scales. Burger's equation
is the simplest type that shows the complicated interplay between
the non-linear steepening and diffusion of a wave. \par
\medskip
The first full solution of the one-dimensional Burgers' equation was
found independently by both Cole [7] and Hopf [12]. Recently, the
solution has been extended to n-dimensions by using group action
on coset bundles [22]. The basic idea of this approach is to determine
coset spaces for a chosen group and for a chain of closed subgroups,
and then to construct a bundle out of these spaces. Next, the action
of the group on the bundle is prescribed and a class of tensor-valued
functions called invariants on the coset bundle is introduced. Because
these group invariants take the form of differential equations,
the method has been used to obtain a tensor version of the
n-dimensional Burgers' equation and to show that its solutions can
be constructed from a linear tensor diffusion equation. It has
also been demonstrated that the procedure gives a nonlinear tensor
equation for a generalized Burgers' equation with a higher than second
order derivative. In this
paper, we examine extensions of the basic class of solutions of the
vector Burgers' equation, which we discovered before we became aware of
 the work of Wolf et al., to more than one dimension and to curvilinear
coordinate systems. The main advantage of our approach is that it
is extremely simple from a mathematical point of view and that the
results we obtain have the potential to be used in a number of
applications
(see Section III for two simple applications). It must be added,
however, that the solutions obtained in this paper represent a
subclass of more general solutions to the tensor Burgers' equation
discussed by Wolf et al. [22].
\medskip
In what follows, we recapitulate part of the historical derivation
presented first by Cole [7] and Hopf [12]. This derivation is central
to our results in Sections II and III.
\medskip
The equation to be solved is
$$ \p u t + u\p u x = \nu {{\partial ^2u }\over
{\partial x^2}} ,\eqno (1)$$
where $\nu $ is a physical constant (viscosity).
A successful transformation uses
$$ u = \p {\phi } x ,\quad   \phi = \phi(x,t).\eqno (2)$$
Substituting into eq. (1) and integrating term-by-term with respect to
x
$$ \p {\phi } t  + {{1}\over {2}} \left (\p {\phi } x \right )^2 = \nu
\p {^2\phi } {x^2} ,\eqno (3)$$

where the function of integration is omitted. Because eq.
(3) is invariant under the transformation
$$ x\rightarrow ax ,\eqno (4a)$$
$$ t\rightarrow a^2t{ },\eqno (4b)$$
(a=constant), as is the diffusion equation,
this suggests that there exists a solution of the form
$$ \phi (x,t) = \phi (\theta (x,t)),\eqno (5)$$
where $\theta $ is a solution of the diffusion equation
$$ \p {\theta } t  = \nu {{\partial ^2\theta  }\over
{\partial x^2}}.\eqno (6)$$
Substitution into eq. (3) while subtracting eq. (6) gives
$$ \left (\ov {d\phi } {d\theta } \right )^2 = 2\nu  \ov {d^2\phi }
{d\theta ^2} ,\eqno (7a)$$
whose solution is
$$ \phi (\theta ) = -2\nu { }\ln { }(\theta - c_1) + c_2,\eqno (7b)$$
so that by choosing $c_1$=0
$$ u(x,t) = {{-2\nu }\over {\theta }}\p {\theta } x .\eqno (8)$$
Quite an interesting literature has developed around applications
of the solutions in pure mathematics [20],
magnetohydrodynamics [9], astrophysics [14,16], and cosmology
[10,15].
Lighthill [18] extended the basic transformation to the case
where the
x coordinate is measured in the frame of reference moving in the same
direction as the wave at the undisturbed speed of sound.
A valuable book on Burgers' equation and related topics was
recently published but is, unfortunately, already out of print [11].
\medskip
In the following, we formally present the derivation of
the solutions to the vector Burgers' equation by using a
generalized Cole-Hopf transformation.
\vskip .1in
{\bf 2. The Vector Extension of Burgers' Equation}
\vskip .1in
We begin with the vector equivalent of Burgers' equation
$$ \p  {\vec u} t  + \vec u\cdot \nabla \vec u = \nu \nabla ^2\vec u.
\eqno (9)$$
In order to evaluate the vector Laplacian we must use
$$ \nabla ^2\vec u = \nabla (\nabla \cdot \vec u) - \nabla \times
\nabla \times \vec u,\eqno (10a)$$
and for the inertial term
$$ \vec u\cdot \nabla \vec u = \nabla \ov  {u^2} 2  - \vec u\times
\nabla \times \vec u.\eqno (10b)$$
Then eq. (9) greatly simplifies as long as
$$\vec u\times \nabla \times \vec u = \nu \nabla \times \nabla \times
\vec u.\eqno (10c)$$
This appears to be an equation that represents quite general flows,
but the only solutions are irrotational. This can be seen by taking
the divergence of both sides of eq. (10c) and using
$$\nabla \cdot (\vec A\times \vec B) = \vec B\cdot \nabla \times
\vec A - \vec A\cdot \nabla \times \vec B.\eqno (10d)$$
In what follows we will assume that the flow field is irrotational
as this is necessary to simplify eq. (9) and to derive the
extension of the Cole-Hopf transformation. The assumption is
satisfied trivially for the one-dimensional Cartesian problem.
$$ \nabla \times \vec u = 0,\eqno (11a)$$
so that
$$ \vec u = \nabla \phi .\eqno (11b)$$
Eq. (11b) is the extension of eq. (2) to three dimensions and,
together with eqs. (10) allows
the simplification of eq. (9) to
$$ \nabla \left [\p {\phi } t  + {{(\nabla \phi )^2}\over {2}} -
\nu \nabla ^2\phi \right ] = 0.\eqno (12)$$
The meaning of eq. (12) is that the quantity enclosed by brackets must
be a function of time, only, because its gradient is zero in all space
so that
$$ \p {\phi } t  + {{(\nabla \phi )^2}\over {2}} -
\nu \nabla ^2\phi  = E(t).\eqno (13)$$
The form of the unknown function E(t) is irrelevant for our discussion
as it cannot affect the velocity. We simply define a new potential
$$ \phi _1 = \phi -\int E(t)dt,$$ where the gradients of $\phi $ and
$\phi _1$ are identical.
\medskip
Before introducing the Cole-Hopf assumption, it is useful to state the
following vector identity:
$$ \nabla \alpha  = \p  {\alpha } {\theta } \nabla \theta ,\eqno (14a)$$
which is true for any orthogonal, curvilinear coordinate system.
Taking the divergence of both sides with $\alpha =\phi $ and using eq.
(14a) again with $\alpha =\p {\phi } {\theta } $ yields
$$ \nabla ^2\phi  = \p {^2\phi } {\theta ^2} \thinspace
(\nabla \theta )^2 +
\p {\phi } {\theta } \thinspace \nabla ^2 \theta .\eqno (14b)$$
We will also need the diffusion equation
$$ \p {\theta } t  = \nu \nabla ^2\theta .\eqno (15)$$
Now we apply eqs. (5), (14b), and (15) to eq. (13) to generate eq.
(7a), again, so that
$$ \vec u = -\ov {2\nu } {\theta } \thinspace \nabla \theta
.\eqno (16)$$
This clearly shows that the Cole-Hopf transformation goes through
for quite general coordinate systems.
Cole [7] originally discussed the vector solution
for the case of cartesian coordinates and he indicated
that it applied to some
special solutions in higher dimensions.
\medskip
The usefullness of eq. (16) can be shown by exploiting the simple
symmetry of the following example.
\vskip .1in
\centerline {\bf 3. Cylindrical Symmetry}
\vskip .1in
Using only a radial velocity,
Burgers' equation in cylindrical coordinates with axisymmetry
may be written as
$$ \p  u t  + u\p  u r  = \nu \left [\ov  1 r \p  { } r \left (
r\p  u r \right ) - \ov  u {r^2} \right ]\eqno (17)$$
Equation (17) can be written as
$$ \p  u t  + u\p  u r  = \nu  \p  { } r \left (\ov  1 r \p  { } r
(ru)\right )\eqno (18)$$
Equation (18) and the related integrals of (18) are what led us to believe
that the Cole-Hopf transformation was quite general in nature.
The particular form of (16) is
$$ u = \ov  {-2\nu } {\theta } \p  {\theta } r \eqno (19)$$
where the diffusion equation is
$$ \p  {\theta } t  = \ov  {\nu } r \p  { } r \left (r\p
{\theta } r \right )\eqno (20)$$
A simple example of the use of this method is to use the initial-value
solution of the diffusion equation (a Dirac delta at t=r=0)
$$ \theta  = \ov  1 {2\nu t} e^{-{{r^2}\over {4\nu t}}} \eqno (21)$$
Then the solution of (17) is found from (19)
$$ u = {{r}\over {t}}\eqno (22)$$

This is the cylindrical version of Burgers' cartesian solution [4],
$u=x/t$, which is the inviscid solution of (1),
(solution 2.1, figure 5 in Benton [3]). Because (22) satisfies (17)
in the limit that $\nu \rightarrow 0$ (which is also the case of zero
acceleration for fluid particles), this solution is the inviscid
limit of the solution of Burgers' equation in cylindrical coordinates.
 Some solutions of (17), such as example B below, approach the
sawtooth limit as $t\rightarrow \infty $; but this is not
surprising as it has been
noted [2,13] that  the
sawtooth shape of dissipation layers is of fundamental importance
in Burgers' representation of turbulence.
\medskip
It is extremely useful to derive solutions of the diffusion equation
which are based on physical boundary conditions on the velocity field.
To this end,
equation (19) may be formally integrated to
$$ \theta (r,t) = k(t)\thinspace \exp \left (-{{1}\over
{2\nu }}\int _0^r u(w,t)\thinspace dw\right )\eqno (23)$$
where
$$ k(t) = \theta (0,t)\eqno (24)$$
The velocity is to be initially specified over a given range so
that the initial value of $\theta $ is given by
$$ \theta (r,0) = \theta _0(r) = k_0\thinspace \exp \left (-{{1}\over
{2\nu }}\int  u_0(w)\thinspace dw\right )\eqno (25)$$
for  $r_0\le r\le R$.
\medskip
The general solution of the diffusion equation, (20),
with azimuthal symmetry, is [19]
$$ \theta (r,t) = \int _0^\infty {\int _0^\infty  \theta _0(r^\prime )
J_0(kr^\prime )J_0(kr)e^ {-\nu k^2t}k\thinspace r^\prime dk}\thinspace
dr^\prime \eqno (26)$$
The integral over k can be done once and for all as it is
a special case of Weber's second exponential
integral as discussed in Watson [21].
$$ \int _0^\infty \exp (-\nu tk^2)J_0(kr)J_0(kr^\prime)k\thinspace
dk = {{1}\over {2\nu t}}\exp \left (-{{r^2+r^{\prime ^2}}\over {4\nu t}
}\right )I_0\left ({{r\thinspace r^\prime }\over {2\nu t}}\right )
\eqno (27)$$
where the modified Bessel function of order zero may be written as
$$ I_0(\alpha r^\prime) = J_0(i\alpha r^\prime )\eqno (28)$$
which is a real function.
Equation (26) may now be written as
$$ \theta (r,t) = \ov  1 {2\nu t} \exp {\left (-\ov  {r^2} {4\nu t}
\right )}
\int _0^{\infty }\theta _0(r^{\prime })\exp {\left
(- \ov  {r^{\prime 2}} {4\nu t} \right )}\thinspace J_0\left (\ov
{ir^{\prime }r} {2\nu t} \right )r^{\prime }dr^{\prime } \eqno (29)$$
\vskip .1in
\centerline {\bf 4. Two Simple Examples}
\vskip .1in
We choose to work two relatively simple examples so as not to obscure
the simple nature of the transform. It is surprisingly easy to pose
apparently straight-forward problems that cannot be readily carried
to completion because of extremely difficult quadratures. We will treat
 the more difficult case of the decay of an initially
constant velocity wind in a future paper.
\medskip
The reader may have noticed by now that Burger's equation does not
specify the density distribution except in a few limits
besides the trivial one of constant density.
\medskip
{\bf A. Incompressible Flow}
\smallskip
The first case is that of incompressible flow where we take a source
of water emanating from the origin of coordinates
and spreading out in a circularly symmetric pattern.
The initial flow is assumed to
exist from very close to r=0 out to R, which is assumed to be larger
than any other length scales in the problem. We then take the limit
as $R\rightarrow \infty $, in the same spirit as other studies of this
initial-value problem on infinite intervals [2].
The speed must decrease as
$r^{-1}$ to conserve mass in an initially steady flow.
In sum,
$$ u(r,0) = u_0\ov  {r_0} r \eqno (30)$$

$$ r_0\le r\le R $$

so that (25) integrates to
$$ \theta _0 = k_0\left (\ov  {r_0} r \right )^a\eqno (31)$$
where
$$ a = \ov  {u_0r_0} {2\nu } \eqno (32)$$
Clearly, $u_0$ may be large at the small value $r_0$ in such a way
that the product $u_0r_0$ is well-defined.
Equation (29) can now be integrated using Hankel's generalization
of (27) as discussed in Watson [21], so that
$$ \theta = k_0r_0^a \thinspace \Gamma \left (1-\ov  a 2
\right )\thinspace
(4\nu t)^{-a/2}\sum _{n=0}^\infty \ov  {\left (\ov  a 2 \right )_n}
{(n!)^2} \left (-\ov  {r^2} {4\nu t} \right )^n\eqno (33)$$
where $\left (\ov  a 2 \right )_n$ is Pochhammer notation,
$$\left (\ov  a 2 \right )_n =  \ov  a 2 \left (\ov  a 2 +1\right )
\left (\ov  a 2 +2\right )....
\left (\ov  a 2 +n-1\right )\eqno (34)$$
and
$$ \left (\ov  a 2 \right )_0 = 1 \eqno (35)$$
The full non-linear solution can now be derived from (19)
and expressed in terms of the confluent hypergeometric
function as
$$ u(r,t) = \ov  a 2 \ov  r t \ov  {M(a/2+1,2,-r^2/4\nu t)} {M(a/2,
1,-r^2/4\nu t)} \eqno (36a)$$
or more clearly as
$$ u(r,t) = \ov  r t \thinspace \ov  {\sum_{n=1}^\infty
(\ov  a 2 )_n\thinspace \ov
n {(n!)^2} \left (-\ov  {r^2} {4\nu t} \right )^{n-1}} {\sum_{n=0}
^\infty \ov  1 {(n!)^2} (\ov  a 2 )_n\left (-\ov  {r^2} {4\nu t}
\right )^n} \eqno (36b)$$

We see that the solution can be written as the product of the inviscid
solution times an infinite sum of powers of the similarity variable
$\ov  {r^2} {4\nu t} $. We note the non-linear effect on higher powers
of the similarity variable through the occurrence of the factor n in
the numerator. The transient non-linearly steepens but diffusion
dominates;
note that the transient is not by itself a solution of Burgers'
equation as we do not have the superposition theorem of linear
theory.
\medskip
Figure 1 was created by non-dimensionalizing equation (17) in the
usual fashion, i.e., $r\rightarrow r/L, t\rightarrow \nu t/l^2,
u\rightarrow u/U$. Choosing $U=u_0$ and $L=r_0$, then a is .5.
This ratio of series is initially divergent and required 100 terms
in double precision on a Sun SPARCstation.
The solutions are presented as isochrones, showing the evolution
of the non-linear solution towards a steady-state. The boundary where
the wind was turned off moves outward while diffusion smooths out
the discontinuity. The wind initially
drops off as 1/r so that the peak in the profile approaches the
starting radius for the source as $t\rightarrow 0$. All curves approach
 the steady-state envelope of 1/r while diffusion flattens the leading
edge. Asymptotically the velocity is zero everywhere.
\medskip
{\bf B. The Decay of An Initially Linearly Increasing Wind}
\medskip
Example two is a wind emanating from the origin, again,
but now the velocity linearly increases in magnitude out to
some radius, $r_0$, chosen to be larger than any other length scale
in the problem. We had in mind the initial theoretical treatments
of radiatively-driven winds [6] from early-type stars, but our
description is too simple to be applied to this problem. This is
because of our assumption of cylindrical rather than spherical
symmetry, and also because other effects are now known to be
important in these winds [8].
\medskip
The driving force is imagined to turn-off
at t=0, and then we calculate the decay of the flow.
In sum
$$ u(r,0) = u_0\ov  r {r_0} = \ov r {t_0} \eqno (37a)$$
$$ 0\le r\le r_0  ;\quad t=0$$
where
$$ t_0 = \ov  {r_0} {u_0} \eqno (37b)$$
is half the time for the wind to flow from the source to the
initial outer boundary. Both $r_0,u_0$ are large, but $t_0$ is not
necessarily big.
Then
$$ \theta _0 = k_0\thinspace e^{-br^2} \eqno (38)$$
$$ b = \ov  {u_0} {4\nu r_0} \eqno (39)$$
Proceeding as before
$$ \theta  = \ov  {k_0} {1+\ov  t {t_0} } \exp \left [-\ov  {r^2}
{4\nu t} \thinspace \ov  1 {1+\ov  {t_0} t } \right ]\eqno (40)$$
$$ u(r,t) = \ov  r {t+t_0} \eqno (41)$$
We note that the solution approaches the Burgers' sawtooth
shape as $t\rightarrow \infty $.
\vskip .1in
\centerline {\bf 5. Conclusions}
\vskip .1in
We have given a relatively simple proof that Burgers' equation can
be solved for curvilinear coordinate systems in three dimensions by
assuming irrotational flow and using equation (16) with $\theta $
derived from solutions of the diffusion equation.
We solved two examples in cylindrical coordinates
with circular symmetry to show how to implement the generalized
solution.
Our goal has been to extend the tools of workers who model the
complicated interplay between non-linear steepening
and viscous diffusion in waves.
\vskip .1in
{\bf Acknowledgments.}
\vskip .1in
ZEM acknowledges partial support of this work by the NSF under grant
no. ATM-9119580. EJS acknowledges partial support of this work
under grant no. NAG 2001 from NASA-Goddard Space Flight Center.
\vskip .1in
\centerline {References}
\vskip .1in

[1] H. Bateman, {\it Some recent researches on the motion of fluids},
Mon. Weather Rev. {\bf 43}, 163, (1915).

[2] E. R. Benton, {\it Solutions illustrating the decay of dissipation
layers in Burgers' nonlinear diffusion equation}, Phys. Fluids,
{\bf 10}, 2113, (1967).

[3] E.R. Benton, {\it A table of solutions of the one-dimensional
Burgers equation}, Q. Appl. Math., {\bf 30}, 195, (1972).

[4] J. M. Burgers, {\it Application of a model system to illustrate
some
points of the statistical theory of free turbulence},
Proc. Roy. Neth. Acad. Sci. Amsterdam {\bf 43}, 1, (1940).

[5] ----, {\it The Non-Linear Diffusion Equation}, D. Reidel Pub. Co.,
Dordrecht-Holland, 1974.

[6] J.I. Castor, D.C. Abbott, and R.I. Klein, {\it Radiatively-driven
winds in Of stars}, Astrophys. J.,
{\bf 195}, 157, (1975).

[7]  J. D. Cole, {\it A quasi-linear parabolic equation in
aerodynamics},
Quart. Appl. Math., {\bf 9}, 225, (1951).

[8] D.B. Friend, and D.C. Abbott, {\it The theory of radiatively-driven
 stellar winds. III. Wind models with finite disk correction and
rotation}, Astrophys. J., {\bf 311}, 701, (1986).

[9] Z. A. Gol'berg, {\it Finite amplitude waves in
magnetohydrodynamics},
 JETP, {\bf 15}, 179, (1962).

[10] S. N. Gurbatov, A. I. Saichev, and S. F. Shandarin,
{\it The large-scale structure of the universe in the frame of the
model of non-linear diffusion},
Mon. Not. R. Astr. Soc., {\bf 236}, 385, (1989).

[11] S.N. Gurbatov, A.N. Malakhov, and A.I.Saichev, {\it Nonlinear
random waves and turbulence in nondispersive media: waves, rays, and
particles}, (Manchester University Press, distributed by St.
Martin's Press, New York, 1991).

[12] E. Hopf, {\it The partial differential equation $u_t+uu_x=\mu
u_{xx}$
}, Comm. Pure and Appl. Math., {\bf 3}, 201, (1950).

[13] D.T. Jeng, R. Foerster, S. Haaland, and W.C. Meecham,
{\it Statistical initial-value problem for Burgers' model equation
of turbulence}, Phys. Fluids, {\bf 7}, 2114, (1966).

[14] J. I. Katz and M. L. Green, {\it A Burgers model of
interstellar dynamics}, Astron. Astrophys.,
{\bf 161}, (1986).

[15] L. A. Kofman and S. F. Shandarin, {\it Theory of adhesion for the
large-scale structure of the universe}, Nature, {\bf 334}
, 129, (1988).

[16] L.A. Kofman and A. C. Raga, {\it Modeling structures of knots in
jet flows with the Burgers equation}, Astrophys. J., {\bf 390}
, 359, (1992).

[17] P. Lagerstrom, J. D. Cole, and L. Trilling, {\it Problems in the
Theory of
Viscous Compressible Fluids}, monograph, California Institute of
Technology, (1949).

[18] M. J. Lighthill, {\it Viscosity effects in sound waves of finite
amplitude}, {\it Surveys in Mechanics}, ed.
Batchelor and Davies,
Cambridge Univ. Press, Cambridge, 1956.

[19] H. Margenau and G.M. Murphy, {\it The Mathematics of
Physics and Chemistry},
Von Nostrand Co., London, 1956.

[20] B. Van Der Pol, {\it On a non-linear partial differential
equation
satisfied by the logarithm of the Jacobian theta-functions, with
arithmetical application}, Proc. Acad. Sci. Amsterdam,
{\bf A13}, 261, (1951).

[21] G.N. Watson, {\it A Treatise on the Theory of Bessel Functions},
 second ed., (Cambridge Univ. Press, Cambridge, 1962, pg. 393-396).

[22] K. B. Wolf, L. Hlavaty, and S. Steinberg, {\it Non-linear
differential
equations as invariants under group action on coset bundles: Burgers
and Korteweg-de Vries equation families}, J. Math. Anal. and Applic.,
{\bf 114}, 340, (1986).

\eject
\centerline {\bf Figure Caption}
The non-linear evolution of a wind with an initial spatial dependence
of 1/r. The isochrones show the decay of the wind, and, especially,
the diffusion at the leading edge where the wind was turned off as
this discontinuity propagates outward.
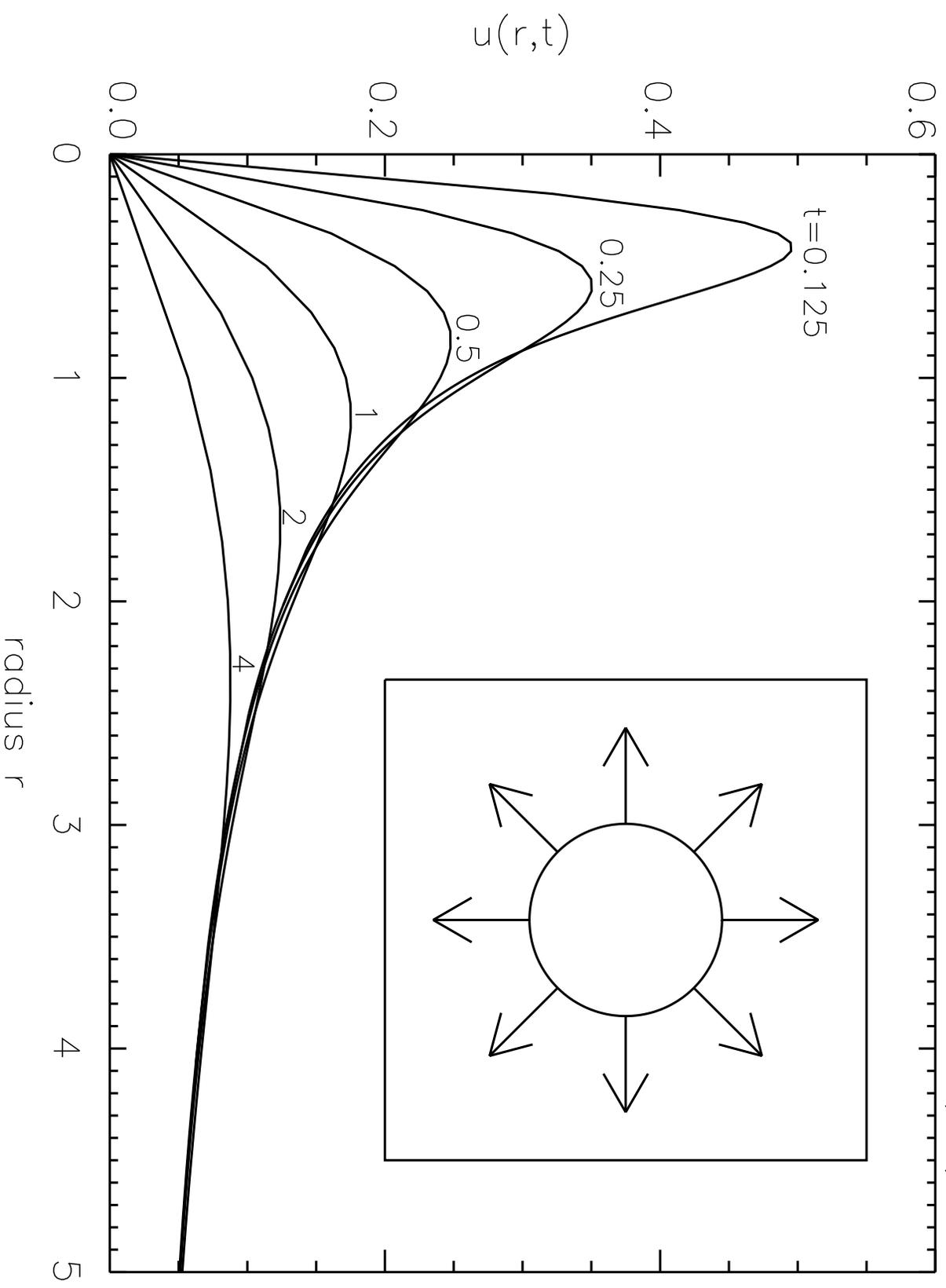
\bye